\documentclass[sigconf]{acmart}
\AtBeginDocument{%
  }

\setcopyright{acmlicensed}
\copyrightyear{2026}
\acmYear{2026}
\acmDOI{XXXXXXX.XXXXXXX}

\acmConference[Conference acronym 'XX]{Make sure to enter the correct conference title from your rights confirmation email}{June 03--05, 2026}{Woodstock, NY}

\acmISBN{978-1-4503-XXXX-X/2026/05}

\usepackage[]{collab}
\usepackage{tcolorbox}
\usepackage{enumitem}

\usepackage{booktabs}
\usepackage{multirow}
\usepackage[table]{xcolor} % 必须添加这个宏包来支持 \rowcolor

\newcommand\alias{\textsc{MASPrism}\xspace}

\collabAuthor{zb}{teal}{Zhuangbin}
\collabAuthor{ly}{purple}{Liuyang}
\collabAuthor{js}{blue}{Junsong}

\begin{document}

% \title[\alias]{\alias: Lightweight Prefill-Only Failure Attribution for Multi-Agent Systems}
\title[\alias]{\alias: Lightweight Failure Attribution for Multi-Agent Systems Using Prefill-Stage Signals}

\author{Yang Liu}
\affiliation{%
  \institution{Sun Yat-sen University}
  \city{Zhuhai}
  \country{China}}
\email{liuy2355@mail2.sysu.edu.cn}
\orcid{0009-0007-0816-0657}

\author{Hongjiang Feng}
\affiliation{%
  \institution{Sun Yat-sen University}
  \city{Zhuhai}
  \country{China}}
\email{fenghj5@mail2.sysu.edu.cn}

\author{Junsong Pu}
\affiliation{%
  \institution{Sun Yat-sen University}
  \city{Zhuhai}
  \country{China}}
\email{pujs@mail2.sysu.edu.cn}

\author{Zhuangbin Chen}
\authornote{Zhuangbin Chen is the corresponding author.}
\affiliation{%
  \institution{Sun Yat-sen University}
  \city{Zhuhai}
  \country{China}}
\email{chenzhb36@mail.sysu.edu.cn}
\orcid{0000-0001-5158-6716}

% \author{Zibin Zheng}
% \affiliation{%
%   \institution{Sun Yat-sen University}
%   \city{Zhuhai}
% \email{zhzibin@mail.sysu.edu.cn}
% \orcid{0000-0001-7872-7718}

% \renewcommand{\shortauthors}{Anonymous Author(s)}

\begin{abstract}
Failure attribution in LLM-based multi-agent systems aims to identify the steps that contribute to a failed execution. 
This task remains difficult because a single execution can contain many agent actions and tool calls, failure evidence can appear many steps after the original mistake, and existing methods often rely on costly agent workflows, replay, or training on synthetic failure logs.
To address these challenges, we propose \alias, a lightweight framework that performs failure attribution using prefill-stage signals from a small language model (SLM).
\alias first extracts token-level negative log-likelihood and attention weights during a prefill pass to identify symptom-like steps and earlier candidate sources, without decoding.
It then reconstructs a focused diagnostic prompt and performs a second prefill pass to rank failure-source candidates. 
Using Qwen3-0.6B as the SLM, \alias achieves the best performance on three of the four evaluated subsets across Who\&When and TRAIL, improving Top-1 accuracy on Who\&When-HC by 33.41\% over the best baseline.
On TRAIL, \alias outperforms strong proprietary LLMs, including Gemini-2.5-Pro, with up to 89.50\% relative improvement.
\alias processes each trace in 2.66 seconds on average, achieving a 6.69$\times$ speedup over the single-pass prompting baseline, with zero output tokens.
% These results suggest that using prefill-stage signals from an SLM is a practical and low-cost way to support failure attribution in long multi-agent execution logs.
These results show that \alias provides an effective and practical framework for failure attribution in long multi-agent execution logs.
\end{abstract}

%%
%% Add CCS concepts here only after the final problem framing is fixed.
%% The author(s) should pick words that accurately describe

\maketitle

\section{Introduction}

LLM-based multi-agent systems increasingly solve complex tasks through coordinated agent actions, inter-agent messages, and tool calls~\cite{DBLP:journals/corr/taubench}.
During execution, these systems produce traces that record the actions taken by different agents and the results returned by external environments~\cite{DBLP:conf/kbse/BouzeniaP25}.
When the final output fails to satisfy the user request, developers need to inspect these traces to identify the steps that contributed to the failure~\cite{DBLP:journals/corr/AgentStepper}.
This attribution process provides the basis for debugging agent workflows, revising prompts, correcting tool-use logic, and improving agent coordination.
Therefore, failure attribution over multi-agent execution traces has become an important problem for maintaining reliable LLM-based multi-agent systems.

Failure attribution over these traces is difficult because the structure of a failed execution often obscures where the failure starts.
First, a trace may contain many agent actions, inter-agent messages, and tool calls~\cite{DBLP:conf/nips/SWE-agent}, creating a large search space for developers to inspect.
Second, the step where a mistake is introduced is often not the step where the failure becomes visible.
An early wrong action may appear locally plausible, but its effect can propagate through later agents and tools before causing an observable error or an incorrect final response.
Third, in realistic debugging settings, developers only have the failed trace and the user request, without access to the correct answer as an oracle for judging each intermediate step.
These properties make failure attribution different from simply detecting an abnormal step: the diagnosis must connect visible symptoms to earlier steps that may have triggered them.

Existing approaches have made failure attribution automated, but they often rely on assumptions that are difficult to satisfy in long-trace and oracle-free debugging settings.
Agent-based methods use a diagnostic agent or workflow to inspect the failed trace and produce an attribution judgment, ranging from direct prompting with a diagnostic agent to multi-step workflows. These methods require decoding over long contexts, which increases latency and token cost for long traces~\cite{DBLP:conf/issta/Xia024APR}.
Replay-based methods collect attribution evidence from repeated executions or trajectory replays to estimate which steps are responsible, which can incur high cost when traces are long or involve tool calls and external environments.
Training-based methods construct annotated or synthetic failure logs to train diagnostic models, which introduces dependence on the coverage and representativeness of the constructed failures.
These approaches provide useful attribution mechanisms, but they often obtain diagnostic evidence through generated reasoning, repeated execution, or task-specific training data, making attribution costly or condition-dependent for long failed traces.

We take a different view: failure attribution can be supported by signals that a language model already computes while reading the failed trace.
In particular, prefill-stage signals from an SLM provide complementary evidence for failure attribution.
% Step-level negative log-likelihood (NLL) highlights symptom-like steps where the failure becomes observable, because explicit errors, inconsistent actions, or abnormal outputs often appear less predictable under the preceding context.
Step-level negative log-likelihood (NLL), computed by averaging token-level NLL within each trace step, highlights symptom-like steps where the failure becomes observable, because explicit errors, inconsistent actions, or abnormal outputs often appear less predictable under the preceding context.
However, these symptom-like steps are not always the earlier sources that triggered the failure.
Attention from these high-NLL steps provides a routing signal that helps prioritize earlier candidate sources.
Our preliminary analysis shows that attention-based ranking from high-NLL steps substantially improves the top-5 coverage of annotated failure-relevant locations, suggesting that these two prefill-stage signals can support symptom-to-source attribution without generating diagnostic text.

\alias operationalizes this observation as a two-stage ranking pipeline that uses only prefill passes of an SLM.
In the Filtering stage, \alias preserves global trace coverage by constructing a truncated trace in which each step is represented by a fixed-length prefix.
It then performs the first prefill pass on this truncated trace, using step-level NLL to identify symptom steps and attention from these symptoms to select earlier candidate sources for diagnosis.
In the Diagnosis stage, \alias reconstructs a focused prompt by restoring the full content of symptom and candidate steps while replacing other steps with brief prefixes and omission markers, and then performs a second prefill pass to rank candidate failure sources.
The framework outputs a ranked list of failure-source candidates with symptom-to-source links, using internal prefill signals rather than decoded explanations or diagnostic judgments.

We evaluate \alias in an oracle-free setting on two complementary benchmarks, Who\&When and TRAIL.
Who\&When evaluates single-step root-cause identification, while TRAIL evaluates localization over multiple annotated error spans in long agent traces.
Using Qwen3-0.6B as the SLM, \alias achieves the best performance on three of the four evaluated subsets, improving Top-1 accuracy on Who\&When-HC from 20.68\% to 27.59\% over the best baseline.
On TRAIL, \alias achieves higher Location Accuracy than Gemini-2.5-Pro, the best baseline that completes both GAIA and SWE-bench, with relative improvements of 8.24\% and 89.50\%, respectively.
At the same time, \alias processes each Who\&When-HC trace in 2.66 seconds on average with 7,066 input tokens and zero output tokens, compared with 17.82 seconds, 17,748 input tokens, and 620 output tokens for A2P.
These results suggest that prefill-stage signals can support competitive attribution accuracy while reducing the cost profile of trace diagnosis.

In summary, this paper makes the following contributions:
\begin{itemize}
    \item We present \alias, an oracle-free failure attribution framework that uses prefill-stage signals from an SLM to diagnose multi-agent execution traces without generated diagnostic text, execution replay, or task-specific training.
    
    \item We design a staged ranking method that retains trace coverage while enabling fine-grained attribution with fewer input tokens. Filtering identifies symptom steps and upstream candidates from a step-level truncated trace, while Diagnosis selectively restores these steps to recover details for final ranking. The framework outputs ranked failure-source candidates with symptom-to-source links, providing inspectable alternatives beyond a single top prediction.

    \item We evaluate \alias on Who\&When and TRAIL, covering both single-step root-cause identification and multi-span error localization. Compared with existing state-of-the-art baselines, the results show higher attribution accuracy on long traces and lower latency and token cost.

\end{itemize}
\section{Background and Motivation}
\label{sec:background}

% ----------------------------------------------------------------
% 2.1  Problem Setting
% ----------------------------------------------------------------

\subsection{Failure Attribution in Multi-agent Systems}
\label{sec:bg:problem}

\subsubsection{Multi-agent System and Its Tracing} A multi-agent system solves complex tasks by dividing work among several language models~\cite{DBLP:conf/iclr/metagpt, li2024survey}. Instead of relying on a single model to process a large amount of information, the system distributes the reasoning process across multiple agents~\cite{DBLP:journals/corr/autogen, DBLP:conf/acl/MapCoder}. These agents communicate through messages and interact with external environments by executing tools~\cite{DBLP:conf/iclr/react,DBLP:journals/corr/MAS}. Through this continuous exchange of data, the system filters intermediate information and refines the working context over time~\cite{DBLP:conf/acl/ChatDev}. The system aggregates outputs from individual agents to gradually construct the final response to the user query~\cite{DBLP:conf/icml/AgentDebate}.
Specifically, the activity record of a multi-agent system can be serialized as a trace $L = \{s_1, s_2, \dots, s_N\}$, where $N$ is the total number of steps. Each step $s_i$ represents a single operation by a specific agent, recording the action it took and the result it received from the environment. However, the final output of a multi-agent system sometimes fails to meet user expectations. 
% When a failure occurs, developers often need to examine the trace to find which agent\ly{and/or which step?} made a mistake so they can optimize the system~\cite{DBLP:journals/tvcg/AgentLens}. 
When a failure occurs, developers often need to examine the trace to identify failure-relevant steps, so that they can optimize the system~\cite{DBLP:journals/tvcg/AgentLens}.

\subsubsection{Problem Statement} To formally describe this diagnostic process, we establish the following definition. \textit{Failure attribution} in multi-agent systems is the process of analyzing a failed trace $L$ to identify the steps that caused the entire system to fail the user query~\cite{DBLP:conf/icml/whowhen, DBLP:journals/corr/TRAIL, DBLP:journals/corr/rethinking, DBLP:conf/iclr/OpenRCA}.
% most responsible for the observed failure.
% rank a small set of steps that most likely caused the observed failure.
% most responsible for the observed failure.
% \zb{the earliest ones?}\ly{I suggest not baking the earliest-step assumption into the general definition, because in Who\&When it is a dataset-specific annotation rule for the decisive error, whereas our paper needs a broader task definition that also remains compatible with TRAIL and our ranking-based formulation.}
% identify a subset of steps $\{s_i\}$ \ly{steps most responsible for the observed failure}\ly{align with the methodology} that caused the entire system to fail the user request. 
Automating this process is challenging. Agents in these systems do not operate in isolation~\cite{DBLP:journals/corr/autogen}. The output of one agent's action often serves as the input for another, creating long and intricate causal chains throughout the trace~\cite{DBLP:journals/corr/TRAIL}. An error made in an early step can therefore propagate through many subsequent actions before the final output reveals a failure~\cite{DBLP:conf/icml/whowhen, DBLP:conf/iclr/react}. This makes tracing the failure back to its original source a difficult task.

\subsubsection{Limitations of Existing Approaches.}
\label{sec:bg:prev_limit}

Recent studies attempt to automate this failure attribution process through several directions.
% We refer to the first direction as \textit{LLM-based prompting for failure attribution}, where researchers ask an LLM to identify the root cause from the failed trace through direct prompting strategies or multi-step diagnostic workflows~\cite{DBLP:conf/icml/whowhen, DBLP:journals/corr/a2p, ECHO, zhu2025raffles}.
% However, to accurately diagnose a failure, the diagnostic model or workflow must thoroughly analyze the long context and the complex interactions between the original agents.
% To achieve this, such methods often require expensive language model calls, especially when they repeatedly review the trace.
% This process makes the financial and computational cost of analyzing a single trace high.
% Since real-world applications generate a large number of failed traces, this high cost can make LLM-based prompting difficult to deploy at scale.
We refer to the first direction as \textit{agent-based} failure attribution, where researchers design a diagnostic agent or workflow to reason over the failed trace $L$ and identify the root cause~\cite{DBLP:conf/icml/whowhen, DBLP:journals/corr/a2p, ECHO, zhu2025raffles}.
Such methods range from direct prompting with a single diagnostic agent to multi-step workflows that repeatedly review the trace or refine intermediate judgments~\cite{DBLP:journals/pacmse/autofl}.
However, to accurately diagnose a failure, the diagnostic agent or workflow~\cite{DBLP:journals/pacmse/agentless, DBLP:conf/icse/RepairAgent} must thoroughly analyze the long context and the complex interactions between the original agents.
This process often requires expensive language model calls, which makes the financial and computational cost of analyzing a single trace high.
Since real-world applications generate a large number of failed traces, this high cost can make agent-based attribution difficult to deploy at scale.

Another direction is \textit{replay-based failure attribution}, where researchers rerun or perturb executions to estimate which steps are responsible for the failure~\cite{DBLP:journals/corr/famas}.
This reliance on repeated replays can incur high attribution cost, especially for long agent traces~\cite{DBLP:journals/corr/chief}.

We refer to the next direction as \textit{training-based failure attribution}, where researchers inject artificial errors into system executions to create synthetic datasets with known root causes and train a diagnostic model~\cite{zhang2026agentracer,kong2026aegis}.
The problem with this approach is that fault injection may capture only a limited set of failure patterns~\cite{DBLP:journals/corr/TRAIL,DBLP:journals/corr/rethinking}.
In the real world, multi-agent systems fail in many different ways~\cite{mast, DBLP:journals/corr/jia2026masfire}, so models trained on synthetic failures may face generalization risks when applied to unseen agent logs~\cite{DBLP:journals/corr/chief}.

Therefore, to make automated failure attribution more practical, we need a new approach.
% \zb{do not talk too much about the industrial side}
Such a method must be computationally efficient enough to handle the large volume of traces generated by real-world systems, by reducing reliance on costly agent workflows~\cite{DBLP:conf/issta/AutoCodeRover} and repeated execution replay.
It should also avoid depending on synthetic error injection to better support diagnosis for diverse and unpredictable failures in new multi-agent systems.

\subsection{Language Model Prefill Stage}
\label{sec:bg:prefill}

Instead of designing costly agent workflows, replaying system executions, or training models on synthetic failure logs, we investigate the internal computations of language models.
% We explore whether the data naturally computed during the initial text reading phase might provide helpful signals to diagnose system failures. 
We explore whether the signals naturally computed during the initial text reading phase can support failure diagnosis.
In the following sections, we first explain these mathematical signals. We then present two empirical studies to observe if these signals could help developers locate abnormal actions and trace causal links within system executions.

\subsubsection{Internal Prefill Signals} Language models generally process text in two phases~\cite{kim2025overfill, DBLP:journals/micro/Splitwise}. The prefill stage reads the entire input sequence at once to build an internal representation of the text~\cite{qin2026prefillasaservice}. After that, the decode stage generates the response one token at a time. During the prefill stage, the model calculates internal states that might be helpful to analyze the system trace. One such calculation is the \textit{attention mechanism}~\cite{DBLP:conf/nips/attentionAlluNeed}. The attention weight $\alpha_{i,j}$ between a current token $i$ and a preceding token $j$ is computed as $\alpha_{i,j} = \text{softmax}(\frac{q_i \cdot k_j}{\sqrt{d}})$~\cite{DBLP:conf/nips/attentionAlluNeed}, where $q_i$ is the query vector of token $i$, $k_j$ is the key vector of token $j$, and $d$ represents the vector dimension. 
This operation determines how much information token $i$ extracts from token $j$. 
% Since these weights directly measure the reliance of later tokens on earlier ones, they can quantify the relevance between different events in the execution trace. 
% This information can then assist in identifying the underlying relevant relationships.
Since these weights indicate how later tokens attend to earlier ones, they can provide a signal for estimating the relevance between different events in the execution trace~\cite{DBLP:conf/emnlp/WiegreffeP19AisnotnotExpla, DBLP:conf/naacl/JainW19AisnotExpla}.
This information can assist in identifying relevant relationships.
% \ly{this claim is so strong. Consider changing to: This information can then assist in routing from later symptom steps to earlier responsible steps.}

Another useful signal from the prefill stage is the \textit{negative log-likelihood} (NLL)~\cite{radford2019language}. For a specific token $x_i$ appearing after a sequence of previous tokens $x_{<i}$, this value is computed as $\text{NLL}(x_i) = -\log P(x_i | x_{<i})$. This equation calculates the negative logarithm of the probability that the model assigns to the actual token. When a token has a high negative log-likelihood, it indicates that the model finds the token highly unlikely to appear given the context. Therefore, this calculation might offer a natural way to identify abnormal actions or surprising errors within the system trace.
% What is prefill stages? What is the outcomes? (NLL, Attention)

% NLL, Attention, 

% 核心是在数据集上 计算 Attention 和 NLL 两组指标，然后发现 我们的问题（failure Attribution）和这两组指标是 **相关** 的

% NLL -> Filtering, Syndrome Discovery;
% Attention -> Casual Chain Construction.

\subsubsection{Empirical Observation of System Anomalies}
\label{sec:bg:obser1}

To analyze system traces at the action level, we aggregate token-level NLL within each step by averaging token scores over the step. We apply this aggregated signal to the Who\&When~\cite{DBLP:conf/icml/whowhen} and TRAIL~\cite{DBLP:journals/corr/TRAIL} benchmarks, which provide failed multi-agent traces with annotated failure-relevant locations. In Who\&When, the annotation is a single root-cause step, whereas in TRAIL it is a set of error spans. For each trace, we rank steps by this aggregated NLL signal in descending order. 
Since this preliminary analysis computes prefill signals over full traces, we report results only on traces that can be processed within the model's context and memory budget. 
The SWE-bench subset of TRAIL is excluded from this observation because its traces are too long for this full-trace analysis.

\noindent\textbf{Observation 1: Higher average token-level NLL tends to highlight symptom-like steps where failure becomes observable.}
As shown by the bars for direct per-step average NLL ranking in Fig.~\ref{fig:empirical_routing}, among analyzable traces, direct ranking by this signal places the annotated root-cause step within the top-5 in 26.67\% of cases on the Hand-Crafted (HC) subset of Who\&When and 53.60\% on the Algorithm-Generated (AG) subset.
These modest hit rates indicate that NLL alone is not sufficient for directly identifying the earlier root-cause step in Who\&When.
On the analyzable GAIA subset of TRAIL, where annotations mark error spans rather than a single decisive source step, at least one annotated error span appears within the top-5 in 87.50\% of cases.

This pattern likely arises because an initial mistake may appear locally plausible when it first occurs.
As the execution continues, its downstream effects can accumulate and become more visible in later steps.
This suggests that token-level surprise, when averaged within each step, is useful for identifying symptom-like steps where failure becomes observable, but may be insufficient for tracing the failure back to its earlier source.

\subsubsection{Empirical Motivation: Local Anomaly is Not Enough}
\label{sec:bg:motivation}

We examine whether attention from high-NLL steps tends to point toward earlier steps that are relevant to the failure. We take the top-ranked NLL steps as symptom-like steps. For each earlier step, we aggregate the attention it receives from these high-NLL steps. We then rank earlier steps by the aggregated attention signal and test whether annotated root-cause steps or error spans move closer to the top.

\begin{figure}[htbp]
  \centering
  \includegraphics[width=\columnwidth]{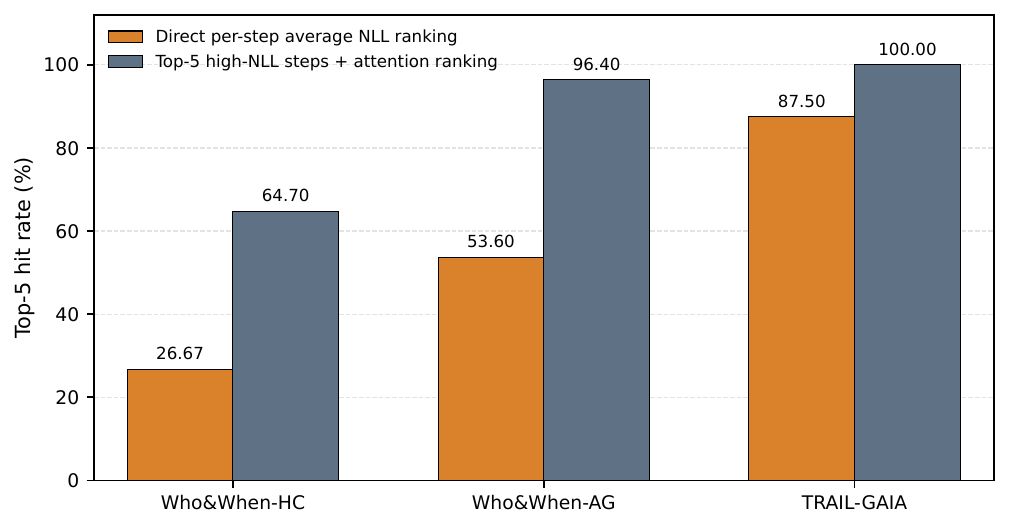}
  \caption{Top-5 Hit Rates of Direct NLL Ranking and Attention-Based Ranking from High-NLL Steps}
  \label{fig:empirical_routing}
\end{figure}

\noindent\textbf{Observation 2: Attention from high-NLL steps helps route diagnosis toward earlier failure sources.}
As shown by the attention-ranking bars in Fig.~\ref{fig:empirical_routing}, ranking earlier steps by the attention they receive from the top-5 high-NLL steps increases the Top-5 hit rate to 64.70\% on Who\&When-HC, 96.40\% on Who\&When-AG, and 100.00\% on the analyzable GAIA subset of TRAIL.

Taken together, these observations suggest that prefill-time signals provide two complementary cues. Average token-level NLL within each step helps identify where a failure becomes observable, while attention from these high-NLL steps helps route the search toward earlier candidate sources. This symptom-to-source use of prefill signals motivates the attribution design described next.

\section{Methodology}

\subsection{Overview} 

\begin{figure*}[t]
    \centering
    \includegraphics[width=\textwidth]{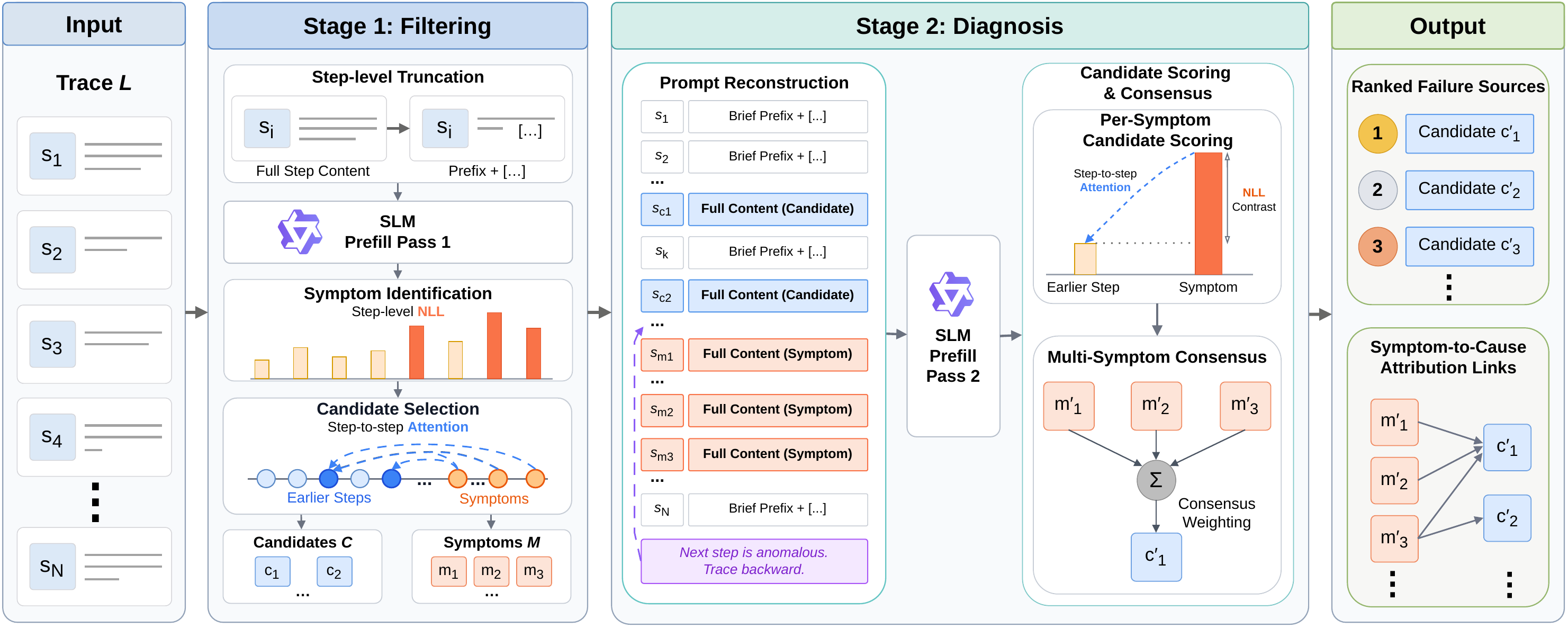}
    \caption{Overview of the \alias\ Framework}
    \label{fig:overview}
\end{figure*}

\alias addresses failure attribution in long multi-agent traces, where each trace consists of a sequence of agent steps. Given such a trace, the pipeline aggregates LLM-internal signals from the prefill pass output, namely NLL and attention weights, at the step level. 
It uses these signals to screen and rank the earlier steps that likely caused the failure and link them to the symptom steps where the errors manifest.
\alias performs this process without the decode phase, relying on two prefill passes of the same SLM.
The pipeline carries out this process in two stages, as illustrated in Fig.~\ref{fig:overview}.

\textit{\textbf{Filtering}} narrows the search space for \textit{Diagnosis} by identifying symptom steps and earlier candidate failure-source steps. 
It reconstructs the trace into a compact prompt by retaining only a fixed-length prefix of each step, which removes redundant details while preserving high-level behavior. From the prefill pass output of this prompt, the framework aggregates NLL at the step level to identify symptom steps and uses attention weights to select the earlier steps that likely caused the failure.

\textbf{\textit{Diagnosis}} performs the final attribution by focusing on the regions identified during \textit{Filtering}. It first rebuilds the trace into a prompt where the symptom and candidate steps are restored to their full content, while all other steps are reduced to brief prefixes. Because these key steps retain their full content, the NLL and attention signals recomputed from a second prefill pass carry richer information than those from \textit{Filtering}. 
For each symptom, \textit{Diagnosis} scores each earlier step by combining the attention the symptom places on it with the NLL contrast between the two steps.
It then aggregates and ranks these scores across all symptoms to produce a final ordered list of candidate failure sources, linking each candidate to its associated symptoms.

\subsection{Filtering: Symptom-Guided Candidate Screening}
The Filtering stage identifies a small number of steps as symptoms and candidate failure sources to narrow the search space for Diagnosis. To ensure the entire trace fits within the model's context window, this stage reconstructs the sequence into a compact prompt by retaining only a fixed-length prefix of each step. This approach removes redundant details while preserving the high-level behavior of every step, enabling the model to observe the global execution flow. By analyzing internal signals from a single prefill pass, the framework uses negative log-likelihood (NLL) to locate symptom steps and attention weights to identify the earlier steps that these symptoms attend to. These steps form the symptom set $\mathcal{M}$ and the candidate set $\mathcal{C}$, which define the scope of the detailed analysis performed in the subsequent Diagnosis stage (Sec.~\ref{sec:diagnosis}).

\subsubsection{Step-level Truncation}

Filtering constructs a truncated version $\tilde{L}$ of the original trace $L = \{s_1, \dots, s_N\}$ to ensure the entire sequence fits within the model's context window. Each step $s_i$ is allocated an equal token budget, and any content exceeding this limit is replaced with an omission marker \texttt{[...]}, resulting in a truncated step $\tilde{s}_i$. This approach prioritizes the breadth of the trace over the depth of individual steps, ensuring that the model can access the entire execution history in a single prefill pass. If a long trace is truncated from one end to fit the context limit, the model often loses either the early steps that contain the root causes or the late steps where symptoms manifest. By retaining a prefix for every step, the framework preserves the global causal chain while avoiding the computational cost of generating summaries for each step. This method relies on the observation that the beginning of an agent's step usually contains its core intent or action, which is often sufficient for preserving the high-level behavior of steps~\cite{DBLP:conf/iclr/react}.

A system prompt (Fig.~\ref{fig:filtering_prompt}) is prepended to inform the model that each step has been truncated and that omission markers represent omitted content. This brief instruction clarifies the formatting, helping the model process the truncated sequence without misinterpreting missing text as errors in the agent's logic~\cite{DBLP:conf/nips/LMFew-ShotLearners}.

\begin{figure}[htbp]
    \centering
\begin{tcolorbox}[title=\textbf{Filtering Prompt}, colback=gray!5!white, colframe=gray!75!black, fonttitle=\bfseries, boxrule=0.5pt, arc=2pt, left=1pt, right=1pt, top=4pt, bottom=4pt]
    \footnotesize

\textbf{[System]} This trace has been truncated. Each step shows a bounded prefix of key content. \texttt{[...]} marks omitted text. Focus on error patterns and causal chains. Omissions are expected.

\end{tcolorbox}
\caption{Prompt Prepended to the Truncated Trace in the Filtering Stage}
\label{fig:filtering_prompt}
\end{figure}

\subsubsection{Step-level Signal Extraction}
\label{sec:step_signal_extraction}

Using the truncated trace $\tilde{L}$, Filtering extracts two internal signals during a single prefill pass to identify the symptom set $\mathcal{M}$ and the candidate set $\mathcal{C}$, following the motivation in Sec.~\ref{sec:bg:motivation}.
The first signal, step-level negative log-likelihood (NLL), helps locate the symptom steps where the failure becomes visible. The second signal, step-to-step attention, measures how strongly these symptoms attend to earlier steps, which helps select the upstream candidate sources.

The step-level NLL for a truncated step $\tilde{s}_i$ is defined as the average token-level negative log-likelihood within that step. Specifically, we compute $N_i = \frac{1}{|T_i|} \sum_{t \in T_i} \ell_t$, where $T_i$ is the set of tokens in $\tilde{s}_i$ and $\ell_t = -\log p(x_t \mid x_{<t})$ represents the negative log-likelihood of token $t$ conditioned on its preceding context. A higher step-level NLL indicates that the step is less predictable to the model, serving as a symptom signal that often coincides with anomalous or error-related behavior, as observed in Sec.~\ref{sec:bg:obser1}.

The second signal is step-to-step attention, which aggregates the token-level weights defined in Sec.~\ref{sec:bg:prefill} to estimate how strongly one step attends to another. 
For each token pair, we first average its attention weights uniformly across all heads and selected layers. 
Let $\bar{\alpha}_{t_i,t_j}$ denote this averaged attention weight from token $t_i$ in a truncated step $\tilde{s}_i$ to token $t_j$ in an earlier truncated step $\tilde{s}_j$. 
The step-to-step attention is then defined as:
\begin{equation}
A_{i \rightarrow j} =
\frac{1}{|T_i|}
\sum_{t_i \in T_i}
\sum_{t_j \in T_j}
\bar{\alpha}_{t_i,t_j}.
\end{equation}
This normalization yields the average attention mass that each token in $\tilde{s}_i$ assigns to the earlier step $\tilde{s}_j$.

\subsubsection{Symptom Identification}
\label{sec:symptom_identification}

Symptom identification produces a symptom set $\mathcal{M}$ by locating steps where the failure becomes visible to the model. This process builds on the principle that NLL measures the degree to which the model finds a specific sequence of tokens unlikely given the preceding context. Because anomalous behaviors or explicit error messages typically represent surprising events that deviate from standard execution, they often result in elevated NLL scores during the prefill pass. Identifying these points of high model surprise helps the framework locate symptoms that may be distributed across multiple steps of the trace.

To rank potential symptoms, the framework evaluates the NLL scores and applies a post-processing refinement to prioritize steps containing explicit failure markers such as \texttt{error} or \texttt{exception}. This prioritization helps move steps with clear indicators of failure to the front of the ranking by combining internal model signals with observable lexical evidence. The framework then selects the top-ranked steps to form the symptom set $\mathcal{M}$. 
The size of this set is proportional to the total number of steps $N$, with ratios specified in Sec.~\ref{sec:implementation_details}. 
This design is intended to provide sufficient coverage of the trace while avoiding an excessive number of symptoms that might otherwise complicate the subsequent analysis.

\subsubsection{Candidate Selection}
Candidate selection identifies potential failure sources by aggregating the attention weights from the symptom set $\mathcal{M}$ to all earlier steps. 
This approach uses attention weights as a routing signal, since they indicate how later tokens draw information from earlier tokens during the prefill pass. 
Attention from symptom steps provides an empirical cue for prioritizing earlier steps, as shown in Sec.~\ref{sec:bg:motivation}.
For each symptom $m \in \mathcal{M}$, the framework examines the step-to-step attention values $A_{m \rightarrow k}$ for all earlier steps $k < m$. 
To identify steps that are consistently linked to the observed failure, the framework computes a global candidate score $H_k$ by summing these weights across all symptoms:
\begin{equation}
H_k = \sum_{m \in \mathcal{M},\, k < m} A_{m \rightarrow k}.
\end{equation}
This aggregation highlights steps that are frequently attended to by multiple symptoms, suggesting they are likely to contain information related to the root cause. The top-$K$ steps ranked by $H_k$ form the candidate set $\mathcal{C}$.
% , where we set $K=5$ in our experiments
% where the value of $K$ is specified in the experimental setup. 
As shown by the empirical evidence in Sec.~\ref{sec:bg:motivation}, this process supports a high probability that the selected steps include the actual failure source by tracing the connections between visible symptoms and earlier steps.
Together, the symptom set $\mathcal{M}$ and the candidate set $\mathcal{C}$ define the relevant steps that are preserved in full detail for the subsequent diagnosis stage.

\subsection{Diagnosis: Multi-Symptom Failure-Source Ranking}
\label{sec:diagnosis}

The Diagnosis stage performs the second prefill pass to evaluate the candidate steps and rank candidate failure sources.
Instead of operating on the uniformly truncated trace, this stage first reconstructs the prompt by restoring the full content of the symptom and candidate steps identified during Filtering.
These steps determine which parts of the trace are preserved in detail for the second prefill pass.
Diagnosis then recomputes NLL and step-to-step attention on the reconstructed prompt, re-identifies symptom steps under the richer context, scores candidate steps for each updated symptom, and applies multi-symptom consensus to produce the final ranking.

\subsubsection{Prompt Reconstruction}

To build this new prompt, the framework constructs a modified trace $L'$ by restoring the full content of the steps in the symptom set $\mathcal{M}$ and the candidate set $\mathcal{C}$ identified during Filtering. All other steps are reduced to a brief prefix followed by an omission marker \texttt{[...]}. This process helps ensure that the relevant steps are available for detailed analysis while the overall sequence remains within the context window of the model. 

The final prompt for this second prefill pass consists of a system prompt and the modified trace $L'$. As shown in Fig.~\ref{fig:diagnosis_prompt}, the system prompt informs the model that the trace has been reconstructed and that omissions are intentional. To further assist the model, the framework inserts an instructional cue into $L'$ immediately before the earliest step of the symptom set $\mathcal{M}$~\cite{DBLP:journals/tacl/lost}. This enhancement frames the surrounding context as a task of tracing the failure back to its source. The reconstruction preserves the original step ordering to maintain the temporal logic of the execution.

\begin{figure}[htbp]
    \centering
\begin{tcolorbox}[title=\textbf{Diagnosis Prompt}, colback=gray!5!white, colframe=gray!75!black, fonttitle=\bfseries, boxrule=0.5pt, arc=2pt, left=1pt, right=1pt, top=4pt, bottom=4pt]
    \footnotesize

\textbf{[System]} This execution trace has been reconstructed. Key steps retain full content; other steps are compressed with \texttt{[...]} placeholders. Analyze which earlier step or location is most responsible for the observed failure.

\medskip
\textit{[Inserted before the earliest symptom step]}

\texttt{[Note]}: The next step shows anomalous behavior. Trace backward to identify which earlier step caused it.

\end{tcolorbox}
\caption{Prompt Prepended to the Reconstructed Trace in the Diagnosis Stage}
\label{fig:diagnosis_prompt}
\end{figure}

\subsubsection{Signal Recomputation}

During the second prefill pass, the framework recomputes the step-level NLL and step-to-step attention on the reconstructed prompt with restored key steps to extract richer diagnostic information.
It calculates $N_i$ and $A_{i \rightarrow j}$ for the steps in the modified trace $L'$ using the same definitions established in Sec.~\ref{sec:step_signal_extraction}. Because the candidate and symptom steps now retain their full content, these recomputed signals capture detailed execution logic and error manifestations that were previously omitted. The framework then reapplies the symptom identification process in Sec.~\ref{sec:symptom_identification} to the new NLL scores to form an updated symptom set $\mathcal{M}'$. These refined signals and the updated symptom set provide the basis for the final candidate scoring.

\subsubsection{Candidate Scoring}

Candidate scoring evaluates the potential of each earlier step to be the root cause by combining the step-to-step attention it receives from a symptom with the relative difference in their uncertainty levels. 
For each symptom step $m \in \mathcal{M}'$, the framework assigns a score $s_{k \mid m}$ to every earlier step $k < m$. This score integrates the normalized step-to-step attention with a directional NLL contrast:
\begin{equation}
s_{k \mid m} = \frac{A_{m \rightarrow k}}{\bar{A}_m} \cdot \left( 1 + \max(0,\, N_m - N_k) \right),
\end{equation}
where $A_{m \rightarrow k}$ is the step-to-step attention from symptom $m$ to candidate step $k$, and $\bar{A}_m$ represents the average attention from symptom $m$ to all steps earlier than $m$.

The NLL contrast term $\max(0, N_m - N_k)$ serves as a directional weight that enhances the attention signal by considering the increase in model surprise. While attention identifies general relevance between steps, this term further emphasizes candidates where the uncertainty increases significantly from the candidate to the symptom. This design is based on the observation that in multi-agent systems, a root cause may appear relatively normal to the model, whereas its downstream impact often results in a clear anomaly. 
% By applying a higher weight when the symptom is notably more surprising than the candidate, the framework tends to favor those earlier steps that are more likely to have triggered the failure. 
By increasing the score only when the symptom is more surprising than the candidate, this term favors attention links that follow the expected pattern from a less anomalous earlier step to a more anomalous downstream symptom.
In practice, the $\max(0, \cdot)$ operation is intended to ensure that this term only provides positive enhancement when a clear difference in uncertainty exists, avoiding any penalty when the candidate and symptom steps show similar levels of model surprise.

\subsubsection{Multi-Symptom Consensus}

Diagnosis aggregates signals across multiple symptoms to produce a final ranking of candidate failure sources. Since failures in long traces often manifest through several symptoms, a candidate that is consistently attended to by multiple symptoms is more likely to be the failure source than a candidate linked to one symptom. The framework first computes a base score $\mathrm{Fuse}(k) = \sum_{m \in \mathcal{M}'} s_{k \mid m}$ by summing the individual scores across all symptoms. 
To prioritize candidates that are consistently identified, the framework applies a consensus factor to steps that appear frequently in the top results of different symptoms:
\begin{equation}
\mathrm{Score}(k) = \mathrm{Fuse}(k) \cdot \left(1 + \lambda \cdot |\mathcal{V}_k|\right),
\end{equation}
where $\mathcal{V}_k \subseteq \mathcal{M}'$ is the set of symptoms that rank step $k$ among their top five earlier steps and $\lambda$ is a constant weight. 
% This mechanism rewards candidates that are identified across multiple symptoms, which helps prioritize steps that are more likely to be the actual failure source.
This mechanism rewards candidates that are identified across multiple symptoms, rather than by one symptom alone, which helps prioritize steps that are more likely to be the actual failure source.

The final output is an ordered list of candidate failure sources, where each candidate is associated with the specific symptoms that led to its identification. This mapping provides a rationale for the results by showing how a specific candidate relates to the observed symptoms across the trace. The framework identifies the top-ranked step as the predicted failure source while providing the full list as a set of alternatives for further inspection.

\section{Evaluation}
\label{sec:eval}

We conduct a comprehensive evaluation to demonstrate the effectiveness and efficiency of \alias. Our evaluation is designed to answer the following research questions:

\begin{itemize}[noitemsep,leftmargin=5.5mm]
    \item \textbf{RQ1:} How effective is \alias in performing end-to-end trace diagnosis?
    \item \textbf{RQ2:} How valid is attention from symptom steps as a diagnostic signal for tracing failures to upstream source steps?
    % Signal Validity of Backward Attention.
    % \item \textbf{RQ3:} Is the three-stage diagnosis pipeline necessary for \alias's effectiveness?
    \item \textbf{RQ3:} What are the individual contributions of the filtering and diagnosis stages to the overall performance of \alias?
    \item \textbf{RQ4:} How efficient is \alias, and how robust is its performance across different language models?
\end{itemize}

\subsection{Experiment Settings}
\label{sec:experimental_setting}

\subsubsection{Datasets}
\label{sec:datasets}

We evaluate \alias on two complementary benchmarks, Who\&When~\cite{DBLP:conf/icml/whowhen} and TRAIL~\cite{DBLP:journals/corr/TRAIL}, to assess its diagnostic capability across different trace lengths, task domains, and annotation structures. 
The Who\&When benchmark comprises a handcrafted subset derived from Magentic-One~\cite{DBLP:journals/corr/Magentic1} and an automated subset generated with CaptainAgent~\cite{DBLP:journals/corr/song2024adaptive}.
It provides annotations of the decisive error step, making it suitable for evaluating single-step root-cause identification in failed multi-agent executions.
The handcrafted (HC) subset contains longer execution traces with an average of 51.60 steps, while the automated (AG) subset contains shorter traces with an average of 8.72 steps, allowing us to examine whether the two-stage design of \alias remains effective as the candidate search space changes.
The TRAIL benchmark focuses on long, structured agentic executions collected as OpenTelemetry traces with the agent-compatible OpenInference format~\cite{DBLP:journals/corr/TRAIL}, covering open-world search in GAIA~\cite{DBLP:conf/iclr/GAIA} and software bug fixing in SWE-bench-Lite~\cite{DBLP:conf/iclr/SWEbench}.
Unlike Who\&When, which annotates a single root-cause step for each failed trace, TRAIL provides multiple annotated error spans, with an average of 5.68 annotated errors per trace. These traces are generated by diverse architectures, including hierarchical and single-agent systems. Therefore, TRAIL evaluates whether a method can recover multiple failure-relevant locations from long traces that often exceed the context limits of standard models. This setting complements Who\&When and tests the robustness of our two-stage design under long-context and multi-span localization scenarios.
Throughout all experiments, we follow the w/o $\mathcal{G}$ setting~\cite{DBLP:conf/icml/whowhen}, where the system performs failure attribution without access to the final ground truth of the user query. This configuration appears to be more challenging and particularly valuable for realistic debugging~\cite{zhang2026agentracer}.

\subsubsection{Evaluation Metrics}
\label{sec:metrics}
We utilize benchmark-specific metrics to align with the distinct annotation structures of each dataset. For the Who\&When benchmark, where each failed trace is mapped to a unique root-cause step, we report Top-1 root-cause identification accuracy. 
This metric represents the percentage of traces where the predicted step exactly matches the annotated cause, providing a direct measure of single-step attribution precision. 
% For the TRAIL benchmark, which involves multiple failure-relevant locations per trace, we follow the official TRAIL scorer and report Location Accuracy. 
% This metric is defined as the recall of the annotated error-location set, calculated by the ratio of correctly identified locations to the total number of ground-truth locations. 
For the TRAIL benchmark, where each trace may contain multiple failure-relevant locations, we follow the official TRAIL scorer and report Location Accuracy (Loc. Acc.). 
Loc. Acc. measures recall over unique annotated error locations with exact span-ID matching. 
\alias represents each predicted location with its OpenInference span identifier and submits up to 10 highest-ranked locations to the scorer. 
For context, the original traces contain 30.59 spans per GAIA trace and 33.77 spans per SWE-bench trace on average.
By employing these complementary metrics, we evaluate the framework's capacity for both precise point-of-failure identification and the robust recovery of complex failure spans across long traces.

\subsubsection{Implementation}
\label{sec:implementation_details}

We implement \alias with Qwen3-0.6B and run it on a server equipped with an NVIDIA GeForce RTX 5090 GPU (Driver 580.76.05, PyTorch 2.7.0+cu128). 
We keep the main diagnostic parameters fixed across benchmarks, except for the symptom ratio, which follows the annotation structure of each evaluation task.
The symptom set contains 20\% of steps for Who\&When and 50\% of steps for TRAIL. 
The larger ratio for TRAIL reflects its multi-span annotations, which require broader symptom coverage than the single-step labels in Who\&When.
Candidate selection keeps the top $K=5$ earlier steps, and Diagnosis uses consensus weight $\lambda=0.3$. 
% For attention extraction, we uniformly average attention over all heads and the last 20\% of transformer layers.
For attention extraction, we uniformly average attention over all heads and the last 20\% of transformer layers, since higher layers tend to capture more task-level and contextual relations than lower lexical layers. 
\alias does not use sampling, decoding, or randomized search during attribution. 
Given the same model, prompts, and diagnostic parameters, the attribution scores are computed by fixed prefill passes and deterministic aggregation rules. 
In repeated checks under the same configuration, minor numerical variation, if any, did not change the final rankings or benchmark scores. 
We therefore report the scores directly rather than confidence intervals over stochastic repeated runs.

\subsubsection{Baselines}
\label{sec:baselines}
We compare \alias against a broad spectrum of baselines, including specialized attribution tools and frontier large language models acting as diagnostic judges. 
Specialized baselines include A2P, which utilizes GLM-5.1 for trace analysis, and AgenTracer, which employs a Qwen3-8B model fine-tuned via reinforcement learning. 
% To ensure a consistent basis for the efficiency and resource consumption analysis in RQ4, we re-implement A2P under the w/o $\mathcal{G}$ setting using its original configuration. 
% For other baselines, we incorporate results directly from their respective benchmark publications to maintain a standardized comparison.
To ensure a consistent basis for the efficiency and resource consumption analysis in RQ4, we re-implement A2P under the same oracle-free w/o $\mathcal{G}$ setting using its released implementation and original configuration. 
For other baselines, we use results from peer-reviewed benchmark papers or official benchmark reports to maintain a standardized and reproducible comparison.
For the Who\&When benchmark, we include results for GPT-4o across three judgment strategies, including all-at-once, step-by-step, and binary search. 
For the TRAIL benchmark, we utilize the official results reported in the original study, which evaluates frontier models such as OpenAI o3, Claude-3.7-Sonnet, and Gemini-2.5-Pro through a standardized prompting judge. 
This comparison examines whether \alias can remain competitive with prompted judge models in long-trace settings by using prefill-stage internal signals and a two-stage filtering-diagnosis design to mitigate context-length and token-cost bottlenecks.

\subsection{RQ1: Effectiveness in End-to-End Diagnosis}

% \ly{这段是不是不用提我们对比的是 oracle-free 的，而是放到Experimental Settings？你分析一下：对的}
% 这两段都没有

Table~\ref{tab:rq1_combined_final} presents the results of the effectiveness evaluation. 
The results show that \alias achieves the highest performance in three out of the four evaluated subsets. On the Who\&When dataset, \alias achieves a Top-1 accuracy of 27.59\% in the HC subset. This represents a 33.41\% relative improvement over the state-of-the-art A2P and more than triples the accuracy of the best GPT-4o baseline, which reaches only 8.77\%. This subset is more challenging than the AG subset because it involves significantly longer execution traces, with an average of 51.60 steps compared to 8.72 steps in AG. 
This performance gap suggests that the filtering-diagnosis design helps identify critical steps within long traces by narrowing the search space.
While A2P shows a higher score of 43.65\% on the AG subset, \alias remains competitive at 36.51\% despite using a significantly smaller language model. These results indicate that internal signals from the prefill stage can effectively identify failure sources without the need for generating tokens or relying on larger language models.

On TRAIL, \alias further shows effectiveness on complex traces. It achieves the highest Location Accuracy on both GAIA and SWE-bench. In the GAIA subset, \alias reaches a score of 0.591, outperforming OpenAI o3 and Gemini-2.5-Pro by 10.47\% and 8.24\%, respectively. The advantage is even more apparent on the SWE-bench subset, where \alias achieves a score of 0.451, representing an 89.50\% relative improvement over Gemini-2.5-Pro. 
This result is notable because \alias uses a 0.6B model while achieving higher Loc. Acc. than much larger models in this setting.
On the SWE-bench subset, several judge models fail to produce results because the execution traces exceed their context limits. 
In contrast, \alias remains executable through its two-stage trace reduction design. 
% This stability indicates that the two-stage approach provides the necessary robustness for analyzing extensive execution traces where standard models often fail. 
This stability suggests that the two-stage approach helps improve robustness when analyzing extensive execution traces where standard models often fail.
By performing well in both single-step identification and locating multiple error spans, \alias shows potential to meet various diagnostic requirements in complex multi-agent system traces.

\begin{table}[t]
\centering
\small
\setlength{\tabcolsep}{4pt}
\caption{End-to-End Diagnosis Results on Who\&When and TRAIL. \texttt{CLE} denotes insufficient context length.}
\label{tab:rq1_combined_final}
\begin{tabular}{llcc}
\toprule
% --- Panel A ---
\multicolumn{4}{l}{\textit{Who\&When Dataset (Root Cause Identification \%)}} \\
\midrule
Method & Model & HC Top-1 & AG Top-1 \\
\midrule
\rowcolor[gray]{0.9} \alias & Qwen3-0.6B & \textbf{27.59} & 36.51 \\
All-at-Once & GPT-4o & 3.51 & 13.53 \\
Step-by-Step & GPT-4o & 8.77 & 15.31 \\
Binary Search & GPT-4o & 6.90 & 16.59 \\
A2P & GLM-5.1 & 20.68 & \textbf{43.65} \\
AgenTracer & Qwen3-8B+RL & 20.68 & 37.30 \\
\midrule
\midrule
% --- Panel B ---
\multicolumn{4}{l}{\textit{TRAIL Dataset (Location Accuracy)}} \\
\midrule
Method & Model & GAIA & SWE-bench \\
\midrule
\rowcolor[gray]{0.9} \alias & Qwen3-0.6B & \textbf{0.591} & \textbf{0.451} \\
TRAIL Prompted LLM Judge & OpenAI o3 & 0.535 & \texttt{CLE} \\
TRAIL Prompted LLM Judge & Claude-3.7-Sonnet & 0.204 & \texttt{CLE} \\
TRAIL Prompted LLM Judge & Gemini-2.5-Pro & 0.546 & 0.238 \\
\bottomrule
\end{tabular}
\end{table}

\subsection{RQ2: Signal Validity of Attention from Symptom Steps}

\begin{figure}[t]
\centering
\includegraphics[width=0.98\linewidth]{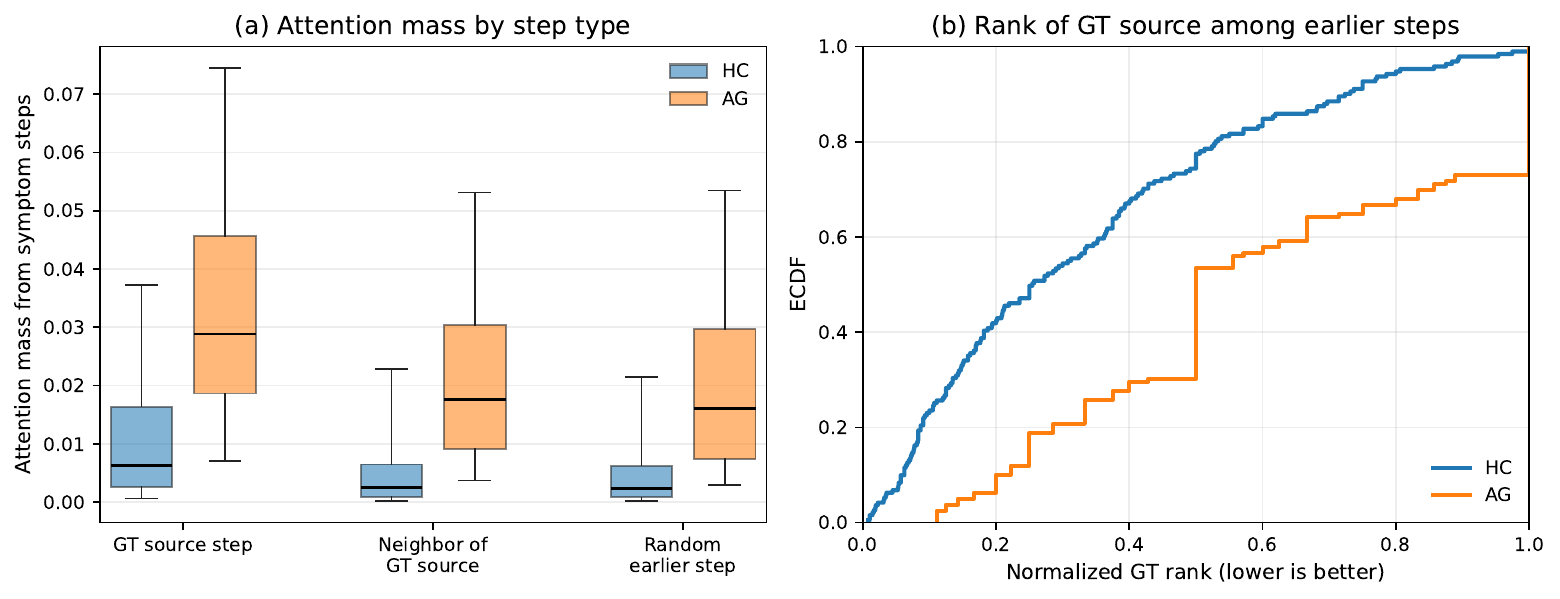}
% \caption{RQ2 Analysis on Who\&When.
% Left: box-plot summaries of attention mass received from symptom steps by GT source steps, neighbors of GT source steps, and random earlier steps.
% Right: ECDF of the normalized rank of the GT source step among earlier candidate steps, where lower ranks indicate stronger prioritization.
% Across both splits, GT source steps tend to receive higher attention mass and appear closer to the top of the attention-based ranking.}
\caption{Attention from Symptom Steps on Who\&When. 
Left: attention mass received by GT source steps, GT-source neighbors, and random earlier steps. 
Right: ECDF of the normalized rank of the GT source step among earlier candidates, where lower is better.}
\label{fig:rq2_attention_main}
\end{figure}

Following the identification of symptoms via NLL, RQ2 evaluates whether attention from symptom steps provides useful guidance for prioritizing upstream failure-relevant steps. As shown in Figure~\ref{fig:rq2_attention_main}(a) and Table~\ref{tab:rq2_attention_validity}, annotated source steps receive higher attention mass from symptom steps than both neighboring source steps and randomly sampled earlier steps across both Who\&When splits. In the Hand-Crafted (HC) split, which contains longer traces with 51.60 steps on average, the difference is statistically significant against both GT-neighbor ($p = 1.90 \times 10^{-9}$) and random earlier-step ($p = 9.82 \times 10^{-9}$) baselines, with a mean difference of $5.24 \times 10^{-3}$. These results indicate that attention from symptom steps provides an empirical routing signal for upstream candidate prioritization.

The same trend appears in the Algorithm-Generated (AG) split. The comparison against GT-neighbor steps remains significant ($p = 3.31 \times 10^{-5}$), while the separation from random earlier steps is weaker but still positive ($p = 4.98 \times 10^{-2}$). This weaker separation is expected because AG traces are shorter on average, which reduces the number of earlier alternatives and makes random earlier steps more likely to be near the annotated source. Figure~\ref{fig:rq2_attention_main}(b) further shows that annotated source steps are often ranked near the top among earlier steps. Together, these results support the use of attention from symptom steps as a practical signal for selecting and ranking upstream candidates in the diagnosis pipeline.

\begin{table}[t]
\centering
\scriptsize
% \caption{RQ2 Paired One-Sided Wilcoxon Signed-Rank Tests. Differences ($\Delta$) are computed per chain as the attention mass of the GT source step minus that of the comparison step.}
\caption{Paired One-Sided Wilcoxon Signed-Rank Tests for Attention Mass. $\Delta$ is the per-chain difference between the GT source step and the comparison step.}
\label{tab:rq2_attention_validity}
\begin{tabular*}{\linewidth}{@{\extracolsep{\fill}}l l c c c c c@{}}
\toprule
Split & Comparison & $n$ & Mean $\Delta$ & Med. $\Delta$ & Win & $p$-value \\
& & & ($10^{-3}$) & ($10^{-3}$) & Rate & \\
\midrule
HC & GT source $>$ GT neighbor & 191 & 5.24 & 1.92 & 0.686 & $1.90\times10^{-9}$ \\
HC & GT source $>$ Random earlier & 189 & 5.15 & 1.81 & 0.640 & $9.82\times10^{-9}$ \\
AG & GT source $>$ GT neighbor & 141 & 7.32 & 5.74 & 0.603 & $3.31\times10^{-5}$ \\
AG & GT source $>$ Random earlier & 86  & 3.74 & 1.99 & 0.558 & $4.98\times10^{-2}$ \\
\bottomrule
\end{tabular*}
\end{table}

\subsection{RQ3: Ablation of the Two-Stage Pipeline}

% The structural ablation results in Table~\ref{tab:rq3_pipeline_ablation} suggest that the synergy between filtering and diagnosis provide complementary benefits, especially on long and complex traces. 
The structural ablation results in Table~\ref{tab:rq3_pipeline_ablation} suggest that filtering and diagnosis provide complementary benefits, especially on long and complex traces.
On the challenging HC split, \alias achieves a Top-1 accuracy of 27.59\%, which is much higher than the 5.17\% observed for the w/o Diagnosis variant. This performance gap indicates that while the first stage successfully narrows the search space, the truncated prefixes used in the compact prompt lack the fine-grained details required for precise diagnosis. 
% Furthermore, the w/o Prompt Restoration variant also remains at 5.17\% on HC traces, which indicates that restoring the full content of candidate steps is a primary driver of accuracy in extended execution traces. 
Furthermore, the w/o Prompt Restoration variant also remains at 5.17\% on HC traces, suggesting that restoring the full content of candidate steps is important for attribution in extended execution traces. 
Although the w/o Diagnosis variant performs reasonably well on simpler AG and TRAIL traces where the execution sequences appear to be shorter, the advantages of the second diagnosis pass become increasingly apparent as the complexity of the agent behavior increases.

The comparison with the w/o Filtering variant, which performs direct single-pass attribution on the raw trace, further demonstrates that the staged narrowing architecture is critical for maintaining operational stability and signal clarity. When attempting to process entire raw traces in a single pass, the Loc. Acc. on the GAIA subset drops significantly to 0.222 compared to the 0.591 achieved by \alias. 
This decline is partly explained by execution failures: the w/o Filtering variant produces 73 out-of-memory failures among the 117 GAIA samples and 11 failures on the HC split. For the remaining executable cases, directly processing raw traces may also weaken signal concentration because the model must distribute attention over substantially longer contexts.
By utilizing the filtering stage as an information bottleneck to prevent context overflow, \alias ensures that the diagnosis stage can operate on fine-grained yet context-efficient prompt structures, which indicates potential for robust end-to-end attribution in multi-agent systems.

\begin{table}[t]
\centering
\footnotesize
\setlength{\tabcolsep}{3pt}
% \caption{RQ3: Pipeline Structure Ablation Across Who\&When (Top-1, \%) and TRAIL (Loc. Acc.)}
\caption{Ablation of the Two-Stage Pipeline on Who\&When and TRAIL}
\label{tab:rq3_pipeline_ablation}
\begin{tabular}{l c c c c}
\toprule
Variant & HC Top-1 (\%) & AG Top-1 (\%) & GAIA Loc. Acc. & SWE Loc. Acc. \\
\midrule
\alias & \textbf{27.59} & \textbf{36.51} & \textbf{0.591} & \textbf{0.451} \\
w/o Diagnosis & 5.17 & 33.33 & 0.511 & 0.381 \\
w/o Prompt Restoration & 5.17 & 35.71 & 0.575 & 0.397 \\
w/o Filtering & 6.90 & 33.33 & 0.222 & 0.384 \\
\bottomrule
\end{tabular}
\vspace{2pt}
\raggedright\footnotesize
% All scores are calculated using the total sample size as the denominator to account for execution failures. The w/o Filtering variant suffers from severe instability, recording 11 failures on HC and 73 OOMs on GAIA.
\end{table}

\subsection{RQ4: Execution Efficiency and Performance Robustness across Language Models}
\label{rq4}

\noindent \textbf{Execution Efficiency.}
We compare \alias with A2P because A2P is the strongest reproducible baseline on Who\&When under the same w/o $\mathcal{G}$ setting.
To make per-trace latency comparable, we run A2P serially, matching the sequential execution of \alias on a single GPU.
As shown in Table~\ref{tab:rq4a_efficiency_hc}, \alias achieves a mean latency of 2.66 seconds per trace on an RTX 5090 GPU, while A2P requires 17.82 seconds through hosted-API calls in our reproduced setting. 
This corresponds to a 6.69$\times$ lower observed end-to-end latency in this deployment setting. 
The latency reduction is consistent with the Filtering-Diagnosis workflow, which extracts NLL and attention during prefill and avoids the costly decode phase~\cite{DBLP:journals/micro/Splitwise, DBLP:conf/osdi/Sarathi-Serve}.
\alias also processes fewer input tokens than A2P (7,066 vs. 17,748), because it uses fixed-length prefixes in Filtering and restores only selected symptom and candidate steps in Diagnosis.
This selective expansion keeps the diagnostic prompt compact, and \alias completes the full attribution process with only two prefill passes and zero output tokens.

\noindent \textbf{Model Robustness.} 
The performance of \alias appears consistent across diverse model architectures and scales, suggesting that internal prefill-stage signals can support failure attribution beyond a single model family. 
As shown in Table~\ref{tab:rq4_b_models}, \alias works across both Qwen and Llama models. 
For localization tasks in GAIA and SWE-bench, increasing the model scale generally improves accuracy, with Llama-3.2-3B achieving the highest SWE-bench score of 0.5116. 
This trend suggests that larger models may capture more detailed diagnostic patterns in long traces.

In contrast, results on Who\&When show that some diagnostic patterns can already be captured by smaller models. 
Qwen3-0.6B achieves a hit count of 16 on the HC subset, which is competitive with its larger 1.7B and 3B counterparts. 
This result indicates that NLL and attention signals remain informative across model sizes, without requiring a large diagnostic model in every setting. 
Such findings support a decoupled deployment strategy, where a lightweight SLM serves as a dedicated diagnostic layer for a more expensive primary agent under constrained computational budgets.

\begin{table}[h]
\centering
\scriptsize
\caption{Execution Efficiency on Who\&When-HC}
\label{tab:rq4a_efficiency_hc}
\begin{tabular*}{\linewidth}{@{\extracolsep{\fill}}l l c c c@{}}
\toprule
Method & Model & Latency (s) & Input Tokens & Output Tokens \\
\midrule
\alias & Qwen3-0.6B & 2.66 & 7,066 & 0 \\
A2P & GLM-5.1 & 17.82 & 17,748 & 620 \\
\bottomrule
\end{tabular*}
\end{table}

\begin{table}[t]
\centering
\scriptsize
\setlength{\tabcolsep}{4pt}
\caption{Diagnostic Performance and Robustness Across Different Model Families and Scales}
\label{tab:rq4_b_models}
\begin{tabular*}{\linewidth}{@{\extracolsep{\fill}}l c c c c c@{}}
\toprule
Model & Params & HC (\#) & AG (\#) & GAIA Loc. Acc. & SWE-bench Loc. Acc. \\
\midrule
Qwen3-0.6B & 0.6B & 16 & 44 & 0.591 & 0.451 \\
Qwen3-1.7B & 1.7B & 13 & 42 & 0.621 & 0.482 \\
Llama-3.2-1B & 1B & 11 & 47 & 0.660 & 0.489 \\
Llama-3.2-3B & 3B & 12 & 50 & 0.643 & 0.512 \\
\bottomrule
\end{tabular*}
\end{table}

\section{Discussion}

\subsection{Practical Deployment Scenario}
The current evaluation validates \alias as a failed-trace attribution method. 
In realistic debugging settings, developers often inspect an execution trace after a failure has been observed, without a correct trajectory or a ground-truth answer for judging each intermediate step. 
In this setting, \alias can serve as a lightweight diagnostic layer for traces submitted for analysis, since it extracts NLL and attention signals through two prefill passes without decoded diagnostic judgments, execution replay, or task-specific training. 
This use case assumes an open-weight SLM whose prefill-stage outputs, including token-level NLL and attention weights, can be accessed during inference. 
Closed API-only services that do not expose these internals are outside the direct deployment target of \alias.
Within this deployment scope, the capability evaluated in this paper is step-level attribution over traces that have already been identified as failed. 
\alias ranks candidate failure-source steps and links them to the symptoms that motivated the ranking. 
In an online setting, these step-level scores could be further aggregated into a trace-level suspiciousness score, where high-scoring traces are escalated for inspection and the Top-$k$ ranked steps guide developers to likely failure sources. 
Because the evaluated benchmarks consist of failed traces, calibrating this trace-level score and controlling false positives on mixed successful and failed runs remain future deployment questions.

\subsection{Using Ranked Candidates in Debugging Workflows}
The ranked output of \alias should be interpreted as a step-level attribution result.
Instead of only producing a single predicted root cause, \alias returns an ordered list of candidate failure-source steps with symptom-to-source links. 
This output better matches practical debugging workflows, where developers may need to inspect several plausible steps before confirming the source of a failure~\cite{DBLP:conf/icse/YangGMH24LLM4testFreeFL}. 
The ranked candidates narrow the inspection scope, while the symptom-to-source links provide an inspectable path from visible failure evidence to earlier suspicious steps. 
This design follows the role separation in the two-stage pipeline: Filtering preserves global trace coverage and selects relevant steps from a compact prompt, while Diagnosis restores these steps and recomputes prefill-stage signals for final ranking. 
Thus, \alias automates the initial attribution step by ranking likely failure-source candidates, helping developers decide where to inspect first.

\subsection{Claim Boundaries and Future Validation}
The internal signals used by \alias should be interpreted as diagnostic routing and ranking evidence, not as formal causal estimators. 
Step-level NLL helps identify where a failure becomes visible to the model, and attention from symptom steps helps prioritize earlier candidate sources. 
A high-ranked candidate therefore indicates a likely failure-source step for early inspection, rather than a confirmed cause, and should be reviewed in context. 
Future validation should study score calibration, false-positive control, and the effect of ranked candidates on manual inspection effort in mixed successful and failed production traces.
\section{Related Work}

\subsection{LLM-based Multi-Agent Systems}
LLM-based multi-agent systems have emerged as an effective paradigm for solving complex tasks by coordinating multiple specialized agents and tool interactions~\cite{DBLP:journals/corr/autogen, DBLP:conf/iclr/metagpt, DBLP:conf/acl/ChatDev}. However, this orchestration introduces complex inter-agent dependencies, where errors can propagate across multiple steps before they become visible in the final outcome. As a result, debugging multi-agent executions becomes more challenging, which makes automated failure attribution an important downstream problem~\cite{mast, DBLP:conf/icml/whowhen}.

\subsection{Failure Attribution in LLM-based Agent Systems}

Failure attribution has recently emerged as an important task for debugging LLM-based agent systems.
Who\&When~\cite{DBLP:conf/icml/whowhen} formulates this task as identifying the agent and decisive step responsible for a failed execution, while TRAIL~\cite{DBLP:journals/corr/TRAIL} broadens the setting to long traces with multiple annotated error spans.

Existing methods obtain attribution evidence through several mechanisms.
Agent-based methods use diagnostic agents or workflows to inspect the trace and predict the responsible step. 
Early methods rely on direct prompting strategies such as all-at-once, step-by-step, and binary-search diagnosis~\cite{DBLP:conf/icml/whowhen}. 
A2P further strengthens the all-at-once setting with an abduct-act-predict scaffold for counterfactual reasoning~\cite{DBLP:journals/corr/a2p}. 
Other methods introduce more structured diagnostic workflows, such as hierarchical context and consensus analysis~\cite{ECHO} or iterative judge-evaluator procedures~\cite{zhu2025raffles}. 
These methods remain dependent on decoded diagnostic judgments over long traces, and multi-step workflows can further increase latency and token cost.
Replay-based methods, such as spectrum-based FAMAS~\cite{DBLP:journals/corr/famas}, estimate failure responsibility through repeated replays and suspiciousness scoring, which can be costly when executions involve tool calls or external environments.
Training-based methods, such as AgenTracer~\cite{zhang2026agentracer} and Aegis~\cite{kong2026aegis}, construct annotated or synthetic failure trajectories to train specialized attribution models, requiring additional data construction and model training.
Beyond method mechanisms, recent benchmark studies suggest that failure attribution may not always reduce to a single deterministic root cause, since multiple steps can be diagnostically informative under complex inter-agent dependencies~\cite{DBLP:journals/corr/rethinking}.

This suggests that single-step prediction may be too restrictive for traces with intricate causal chains.
In contrast, our work treats attribution as a staged ranking task over language-based traces.
It uses prefill-stage signals from an SLM to route from visible symptoms to earlier candidate sources, without relying on costly agent workflows, execution replay, or trained tracers during attribution.

\subsection{Software Fault Localization and Trace Diagnosis}
A common paradigm in software debugging and fault localization is to rank suspicious entities based on failure evidence. A prominent example is traditional fault localization, where spectrum-based techniques such as Tarantula~\cite{DBLP:conf/kbse/tarantula} and Ochiai~\cite{DBLP:journals/jss/ochiai} rank code statements or branches by analyzing execution patterns from test cases. This ranking perspective extends to trace-level diagnosis in complex software systems, where root cause analysis is performed on structured execution traces. For instance, RepTrace~\cite{DBLP:conf/kbse/reptrace} analyzes system call traces through causality analysis, while TraceContrast~\cite{DBLP:conf/icse/tracecontrast} uses sequential pattern mining on distributed telemetry. These methods demonstrate the effectiveness of symptom to source reasoning, but they typically rely on structured execution artifacts such as coverage information, telemetry, or dependency traces~\cite{DBLP:journals/corr/Integrating}. Such artifacts are often insufficient for multi-agent traces, where the crucial causal logic is embedded in free form natural language interactions rather than directly exposed as system native diagnostic dependencies. Our work adopts the same ranking paradigm but operates on language-based agent traces using diagnostic signals extracted during a lightweight model's prefill analysis, instead of depending on system native instrumentation or explicit dependency reconstruction during attribution.

\section{Conclusion}

This paper tackles failure attribution in LLM-based multi-agent systems, where long traces and delayed failure evidence make it difficult to identify failure-relevant steps without an oracle. 
To address this challenge, we presented \alias, a lightweight framework that uses prefill-stage NLL and attention signals from a small language model to diagnose failures without generated diagnostic text, execution replay, or task-specific training. 
\alias identifies symptom-like steps and earlier candidate sources from a compact prompt, then reconstructs a focused prompt for a second prefill pass to rank failure-source candidates. 
Evaluated on Who\&When and TRAIL, \alias achieves the best performance on three of the four evaluated subsets, improves Top-1 accuracy on Who\&When-HC by 33.41\% over the best baseline, and achieves a 6.69$\times$ speedup over the single-pass prompting baseline. 
Overall, these results show that \alias provides an effective and practical framework for failure attribution in long multi-agent execution logs.

\section*{Data Availability}
Code and data are available at
\href{https://github.com/Lycc42/MASPrism}{https://github.com/Lycc42/MASPrism}.

\bibliographystyle{ACM-Reference-Format}
% \bibliography{acmart}
\bibliography{bibliography}

\end{document}